\documentclass[aps,pra,showkeys,twocolumn,superscriptaddress,groupedaddress]{revtex4-1}
\usepackage[dvips]{graphicx}
\usepackage{color,epsfig}
\usepackage{amsmath,amssymb}
\usepackage{graphicx}

\begin{document}

\title{Partial chiral symmetry-breaking as a route to spectrally isolated topological defect states in two-dimensional artificial materials}

\author{Charles Poli}
\affiliation{Department of Physics, Lancaster University, Lancaster, LA1 4YB, United Kingdom}

\author{Henning  Schomerus}\email{h.schomerus@lancaster.ac.uk}
\affiliation{Department of Physics, Lancaster University, Lancaster, LA1 4YB, United Kingdom}

\author{Matthieu Bellec}
\affiliation{Universit\'{e} C\^{o}te d'Azur, CNRS, LPMC, France}

\author{Ulrich Kuhl}
\affiliation{Universit\'{e} C\^{o}te d'Azur, CNRS, LPMC, France}

\author{Fabrice Mortessagne}\email{fabrice.mortessagne@unice.fr}
\affiliation{Universit\'{e} C\^{o}te d'Azur, CNRS, LPMC, France}

\date{\today}

\begin{abstract}
Bipartite quantum systems from the chiral universality classes admit topologically protected zero modes at point defects. However, in two-dimensional systems these states can be difficult to separate from compacton-like localized states that arise from flat bands, formed if the two sublattices support a different number of sites within a unit cell.
Here we identify a natural reduction of chiral symmetry, obtained by coupling sites on the majority sublattice, which gives rise
to spectrally isolated point-defect states, topologically characterized as zero modes supported by the complementary minority sublattice.
We observe these states in a microwave realization of a dimerized Lieb lattice with next-nearest neighbour coupling,
and also demonstrate topological mode selection via sublattice-staggered absorption.
\end{abstract}


\maketitle

\section{Introduction}
Symmetry-protected zero modes are an ubiquitous feature of quantum systems exhibiting nontrivial topological phases. In electronic systems, such phases appear in normal-conducting systems with a chiral  or modified time-reversal symmetry, as well as in superconducting systems where they are due to the charge-conjugation symmetry and may open routes for topological quantum computation \cite{Has10,Qi11,Bee15}. Practical realizations of topological modes also abound in artifical photonic materials \cite{Lu14}, where they enable robust unidirectional transport \cite{Wan08,Wan09,Haf11,Haf13} in analogy to the quantum hall effect and topological insulators \cite{Rag08,Fan12,Kha13,Rec13a,Rec13,Zeu13,mit14}. More generally, zero modes can be created by topological defects and interfaces \cite{Mal09,Kit12,Kra12,Ver13,kei13,pod14}, and display anomalous features \cite{Rud09,Sch13a,Sch13b,Zeu15} that can be exploited, e.g., for topological mode selection \cite{Pol15}.
In all these settings, the topological considerations invoke a combination of symmetry with the dimensionality of the bulk and the defects \cite{Ryu10,Teo10}. In particular, a two-dimensional system constrained only by conventional time-reversal symmetry is topologically trivial \cite{fn1}. The corresponding zero modes, spatially localized at line or point defects, are usually not protected. To obtain nontrivial stationary features, these implementations therefore have to break or modify time-reversal symmetry or rely on system designs that display an additional charge-conjugation or chiral symmetry.

\begin{figure}[t]
\includegraphics[width=\columnwidth]{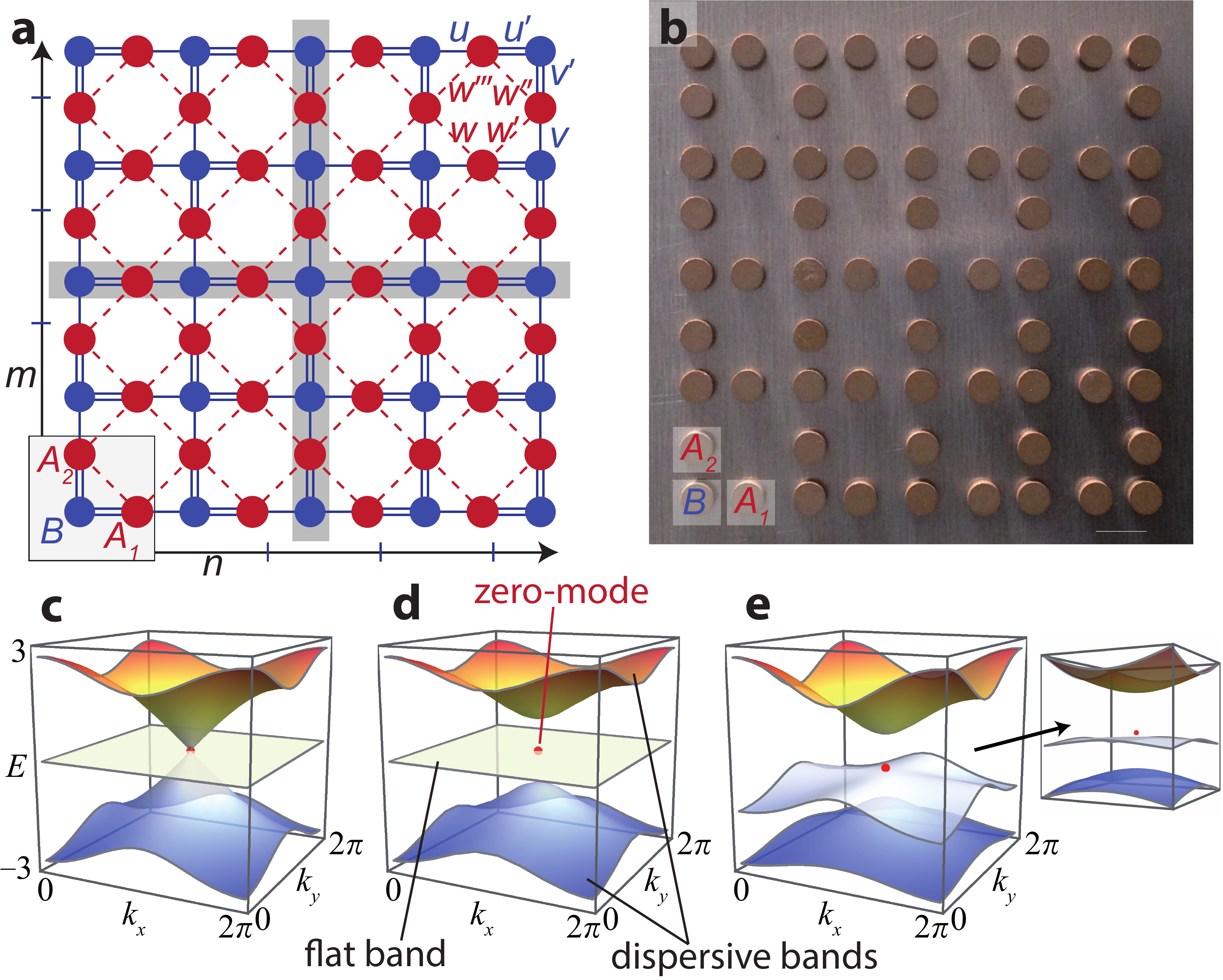}
\caption{\label{fig1} (a) Lieb lattice with dimerized couplings $u$, $u'$, $v$, $v'$, and next-nearest-neighbour couplings $w$, $w'$, $w''$, $w'''$ that reduce the chiral symmetry. This lattice realizes a centered dimerization defect (gray zones). (b) Picture of the experimental microwave realization of the dimerized Lieb lattice. The white bar corresponds to 12 mm. (c-e) Dispersion relation of the infinite system. The localized zero mode (with $\mathrm{Re}\,\mathbf{k}_{-}$ fixed to the M point, red dot) becomes spectrally isolated as one passes from (c) uniform couplings $u=v=u'=v'=1$ over (d) dimerized couplings $u=v=4/3$, $u'=v'=2/3$, gapping out the extended states, to (e) additional next-nearest-neighbour couplings $w=0.4$, $w'=w'''=0.2$, $w''=0.1$, gapping out the flat band. The inset in (e) shows the vicinity of the M point.}
\end{figure}

Chiral symmetries are appealing as they emerge naturally in bipartite lattices \cite{sutherland,Lie89} and place the systems into the same universality class as represented by superconductors coupled to strong topological insulators \cite{Has10,Qi11,Bee15}, meaning that they can support robust zero modes localized at point defects \cite{Teo10}, which can carry a fractional charge \cite{Jac76,Su79,Hou07,Ser08,Ser08a}.
The anomalous behaviour of zero-modes originally identified in the setting of continuum
and lattice field theories \cite{anomaly1,anomaly4,anomaly5} then manifests itself in a finite sublattice polarization \cite{sutherland,Lie89}, while all finite-energy states have an equal weight on both sublattices.
As an intriguing consequence, many bipartite lattices of interest, such as the Lieb lattice presented in Fig.~\ref{fig1}(a), exhibit flat bands of zero modes supported by one of the two sublattices (the majority sublattice, which contains more sites per unit cell than the complementary minority sublattice).
These flat bands provide a competing source of (compacton-like) localized states that are degenerate with any point-defect zero mode  \cite{bergman,Gre10,mur-petit,leykam,Bod14,guzman,Vic15,Muk15},
and also modify constraints on the number of band touchings and Dirac points exhibited by the symmetric dispersive bands of propagating states \cite{sutherland,Lie89,bergman,Gre10,mur-petit}, as illustrated in Fig.~\ref{fig1}(c).

Here, we describe how one can spectrally isolate the point-defect states arising in this setting from the compacton-like states of the flat band. This can be achieved by a well-defined reduction of the chiral symmetry on the majority sublattice, which detunes the compacton-like states due to their finite sublattice polarization (imbalanced weight on the sublattices).
This situation, which we express as a partial chiral symmetry, is realized most naturally in the context of two-dimensional systems, where point-defect states residing on the minority sublattice can be formed when Dirac cones are lifted [see Fig.~\ref{fig1}(d)].
The hybridization of the point-defect states with the compacton-like states can then be prevented by breaking the chiral symmetry on the majority sublattice, which moves the flat band away from the chiral symmetry point in the spectrum, as seen in Fig.~\ref{fig1}(e).
This leaves behind a spatially localized zero mode supported by the minority sublattice, on which the chiral symmetry remains operational.

We first develop a general description of this partial chiral symmetry, and then focus on its concrete experimental implementation in
a dimerized microwave Lieb lattice with next-nearest neighbor couplings [see Fig.~\ref{fig1}(b)].
We observe that the point-defect zero mode displays the expected symmetry-protection against a restricted class of disorder, while  generic disorder only affects it very weakly. As an application, we demonstrate that the finite sublattice polarization of the point-defect state can be exploited for mode selection, which we here achieve by inducing absorption onto the majority sublattice.

\section{Concept and Methods}
\subsection{Partial chiral symmetry}
We first describe our main concept, the formation and spectral isolation of stable defect states against a background of flat bands, in the general context of two-dimensional bipartite lattices \cite{sutherland,Lie89}. Such systems consist of two sublattices (A sites and B sites) that are coupled together to result in the following off-diagonal Bloch Hamiltonian (see the Appendix for further details),
\begin{equation}
h(\mathbf{k})=\left(
\begin{array}{cc}0 & t_{AB}(\mathbf{k}) \\ t_{BA}(\mathbf{k}) & 0 \\ \end{array}\right),\quad
E\varphi=h\varphi,
\quad
\varphi=\left(
\begin{array}{c}\varphi_A \\ \varphi_B \\ \end{array}\right).
\label{eq:tb}
\end{equation}
Here $t_{AB}(\mathbf{k})=t_{BA}^\dagger(\mathbf{k})$ describes the coupling of the A sites to the B sites, whose amplitudes
are collected into vectors $\varphi_A$ and  $\varphi_B$. The coupling term
$t_{AB}(\mathbf{k})$ is thus an $n_A\times n_B$ matrix whose dimensions are given by the count of sublattice sites in the unit cell.
We set $n_A\geq n_B$ and identify the A and B sites with the majority and minority sublattice, respectively.
Generically, there are then $n_A-n_B$ zero modes for any wavevector $\mathbf{k}$,
given by the solutions of $t_{AB}(\mathbf{k})\varphi_A=0$ while $\varphi_B=0$. These dispersionless states
constitute the sublattice-polarized flat bands. They are complemented by $2n_B$ dispersive bands of extended states that occur in pairs of opposite energies, while the Bloch wavefunctions carry equal weight on both sublattices. All of these features are intimately linked to the chiral symmetry, $\tau_z h(\mathbf{k}) \tau_z=-h(\mathbf{k})$, where the Pauli matrix $\tau_z$ acts in sublattice space; in particular, only the zero modes display a finite sublattice polarization $\mathcal{P}=|\varphi_A|^2-|\varphi_B|^2$.

As $\mathbf{k}$ is varied over the Brillouin zone, the dispersive bands can touch at $E=0$, meaning that they cross the flat band. This corresponds to situations where the columns of $t_{AB}(\mathbf{k})$ become linearly dependent of each other.
We focus on scenarios where this happens at discrete Dirac points $\mathbf{k}=\mathbf{k}_p$ ($p$ integer).
If a Dirac point occurs at real $\mathbf{k}$, one finds two additional zero modes, obtained by the degeneracy of the finite-energy states as $\mathbf{k}\to\mathbf{k}_p$. Exploiting that these states are related by $\tau_z$, they can be combined into a mode $\varphi_+$ on the majority sublattice, as well as a mode $\varphi_-$ on the minority sublattice---which then obeys $\varphi_{-,A}=0$, $t_{AB}(\mathbf{k}_p)\varphi_{-,B}=0$. We now aim to spatially confine and spectrally isolate the state $\varphi_{-}$.

In order to create, as a first step, spatially localized variants of this mode, we have to consider situations where the Dirac point is lifted, so that it lies in the complex ${\bf k}$-plane and describes evanescent zero modes. We identify three important features of these modes: (i) they appear in pairs of distinct wave vectors $\mathbf{k}_{\pm}$, (ii) they are supported by opposite sublattices, and (iii) in a finite geometry, only one of them can be compatible with the boundary conditions. In detail, the mode $\varphi_+$ localized on the majority sublattice corresponds to the case of linearly dependent rows of $t_{BA}(\mathbf{k}_{+})$, while the mode $\varphi_-$ localized on the minority sublattice corresponds to linearly dependent rows of $t_{AB}(\mathbf{k}_{-})$. As $t_{BA}(\mathbf{k})=[t_{AB}(\mathbf{k}^*)]^\dagger$, the complex wave vectors $\mathbf{k}_{+}$ and $\mathbf{k}_{-}=\mathbf{k}_{+}^*$ are distinct and describe states that decay into opposite directions. In a finite system, the mode $\varphi_+$ is automatically compatible with the boundary conditions if the system is terminated on the majority sublattice, while for $\varphi_-$ this is the case for termination on the minority sublattice---the sites just beyond the boundary then lie on the sublattice where the amplitude of the mode vanishes. This zero mode is then exponentially localized around a corner of the system,  and retains its finite sublattice polarization $\mathcal{P}=\pm 1$.

To move this localized zero mode to an arbitrary position within the system, we can create a crossing of line defects  that join four regions, in which the mode decays as one moves away from the resulting point defect, as illustrated in Fig.~\ref{fig1}(a). The matching conditions are then automatically met if the interface is formed by the appropriate sublattice.

Recall that this zero mode can still hybridize with the flat band. The flat band can now be gapped out by introducing symmetry-breaking terms $t_{AA}$ (but not $t_{BB}$) into Eq.~\eqref{eq:tb}, leading to a Bloch Hamiltonian of the form
\begin{equation}
h(\mathbf{k})=\left(
\begin{array}{cc}t_{AA}(\mathbf{k}) & t_{AB}(\mathbf{k}) \\ t_{BA}(\mathbf{k}) & 0 \\ \end{array}\right).
\label{eq:semichiral}
\end{equation}
Note that this Bloch Hamiltonian fulfills
\begin{equation}
[\tau_z
h(\mathbf{k})\tau_z]_{BB}=[-h(\mathbf{k})]_{BB},
\end{equation}
which we take as the formal definition of a partial chiral symmetry.
While the final configuration does not exhibit a global chiral symmetry, the symmetry is thus still operational on the minority sublattice, and if the point-defect mode was localized on this sublattice neither its wavefunction nor its energy are affected. This then leaves behind a spectrally isolated and spatially localized zero mode supported by the minority sublattice.

\subsection{Experimental realization for a Lieb lattice}
The regular Lieb lattice \cite{Lie89} consists of a square sublattice B, with additional sites A$_1$ and A$_2$ (forming the A sublattice) placed into the unit cell [gray rectangle in Fig.~\ref{fig1}(a)] so that they subdivide each horizontal or vertical edge. This regular lattice has recently been realized in photonic lattices, allowing to directly observe the compacton-like states \cite{Vic15,Muk15}. Figure~\ref{fig1}(a) shows a dimerized version of the Lieb lattice
(with a central defect obtained as discussed below), and Fig.~\ref{fig1}(b) its experimental implementation. This consists of an array of 65 microwave cylindrical resonators (5 mm height, 8 mm diameter) made of ZrSnTiO (refractive index 6), placed between two metallic plates [top plate not shown in Fig.~\ref{fig1}(b)].

The resonators possess a bare resonance at $\nu_0=6.65$ GHz, which is well isolated within a large frequency interval, and are coupled via the evanescent field, with a distance-dependence that has been characterized in detail in Refs. \cite{Kuh10,Bel13b}.
If we denote by $\Psi_0$ the wavefunction associated to the resonance, the $z$-component of the magnetic field reads  (see \cite{Bel13b}, Eq. (4), for the complete expression of the field) $B_z(r,z) = B_0 \sin \left(\frac{\pi}{h} z\right) \Psi_0(r)$ with
\begin{equation}
\label{eq:Bz}
\Psi_0(r) =
\begin{cases}
 J_0(\gamma_j r) & \text{inside}, \\
 \alpha K_0(\gamma_k r) & \text{outside},
\end{cases}
\end{equation}
where $\Psi_0(0) = 1$. Here $J_0$ and $K_0$ are Bessel functions, $r$ is the distance from the center of the resonator, $\gamma_{j} = \sqrt{\left(\frac{2 \pi \nu_0 n}{c}\right)^2 - \left(\frac{\pi}{h}\right)^2}$, and $\gamma_{k} = \sqrt{\left(\frac{\pi}{h}\right)^2-\left(\frac{2 \pi \nu_0}{c}\right)^2}$ ($n$ denoting the refractive index). Recall that far from the origin, $K_0(x)$ is essentially approximated by $\exp(-x)/\sqrt{x}$. The coupling energy between two adjacent discs depends on the disc separation $d$ and can be described, accordingly to Eq.~\eqref{eq:Bz}, by a modified Bessel function $\left| K_0 ( \gamma_k  d/2) \right|^2$ \cite{Bel13c}. As one infers from in Fig.~\ref{fig1}(a,b), in the Lieb lattice the dominant coupling is between nearest neighbors ($u$, $u'$, $v$, $v'$), but next-nearest neighbor couplings of sites on the A sublattice [represented by $w$, $w'$, $w''$ and $w'''$ in Fig.~\ref{fig1}(a)] are appreciable, while direct couplings on the B sublattice are negligible.

Using a vectorial network analyzer, the reflected signal $S_{11}$ at position $\mathbf{r}_1$ is measured by adequately positioning a movable loop-antenna.  From the amplitude and phase of $S_{11}$ measured over a given frequency range, we have direct access to the local density of states (LDOS)
\begin{equation}
\rho(\mathbf{r}_1, E) \propto \frac{\vert S_{11}(E)\vert^2}{\langle \vert S_{11}\vert^2 \rangle_{E}} \varphi_{11}'(E),
\label{Eq_rho}
\end{equation}
where $\langle\ldots\rangle_E$ indicates an average over the whole accessible frequency spectrum, $\varphi_{11}=\mathrm{Arg}(S_{11})$ is the phase of the reflected signal, and $\varphi_{11}'$ denotes its derivative with respect to the frequency. The density of states (DOS) is obtained by averaging over all positions $\mathbf{r}_1$.
Moreover, our set-up allows us to visualize the wavefunction associated with each eigenfrequency. According to the definition of the local density of states $\rho(\mathbf{r}_1, E) =\sum_n\vert\Psi_n(\mathbf{r}_1)\vert^2\delta(E-E_n)$, the resonance curve exhibits a maximum whose value is related to the intensity of the wavefunction at the specific position $\mathbf{r}_1$. The intensity distribution for the  wavefunction associated with each individual eigenfrequency thus becomes directly accessible.

\begin{figure}[t]
\includegraphics[width=\columnwidth]{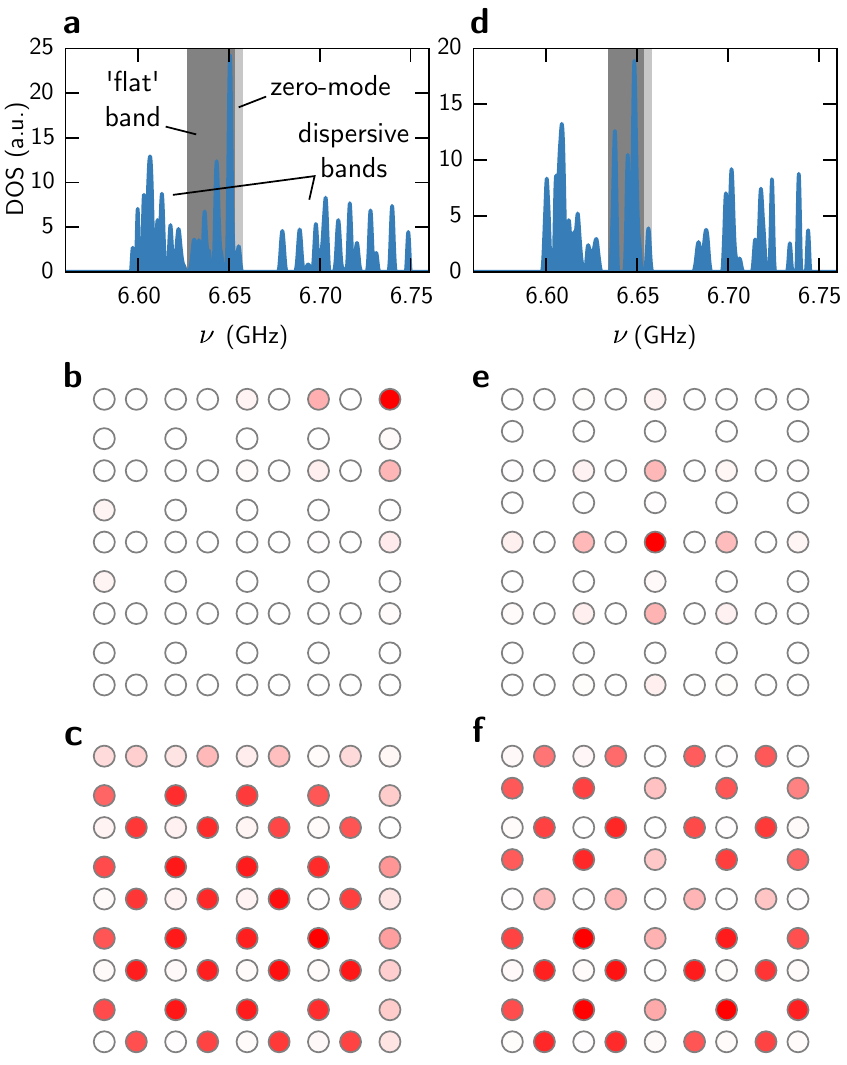}
\caption{\label{fig2} Experimental results for the microwave realization of the partial chiral dimerized Lieb lattice, with a defect state localized in the corner (a-c) or in the center (d-f).
(a,d) Density of states, with the point-defect zero mode shaded in light gray and the flat band shaded in dark grey. (b,e) Spatial distribution of the zero mode. (c,f) Spatial distribution of the flat band.}
\end{figure}

\section{Results}
\subsection{Detection of defect states}

Figures~\ref{fig1}(c-e) show the predicted evolution of the band structure with the couplings, obtained by diagonalizing the corresponding Bloch Hamiltonian (see the Appendix for details).
Figure~\ref{fig1}(c) corresponds to the situation of a regular Lieb lattice with uniform couplings  $u=v=u'=v'=1$ and $w^{(i)}=0$. The band structure consists of a flat band of states supported by the A sublattice and two dispersive bands of extended states touching at a conical Dirac point, positioned at the M point $\mathbf{K}_0=(\pi,\pi)$. As shown on Fig.~\ref{fig1}(d), by introducing dimerized couplings $u=v=4/3$, $u'=v'=2/3$, a gap $2\Delta$, with $\Delta^2=(u-u')^2+(v-v')^2$, opens up between the dispersive bands, while the flat band and the point-defect state remain fixed at zero energy. This corresponds to a virtual Dirac point at
$\mathbf{k}_{\pm}=\mathbf{K}_0\pm i\mathcal{A}$, with $\mathcal{A}=[\ln(u/u'),\ln(v/v')]$.
Additional next-nearest-neighbour couplings $w=0.4$, $w'=w'''=0.2$, $w''=0.1$ make the flat band dispersive and move it to finite energies ($|E|\geq|w+w''-w'-w'''|\equiv \Delta_0$), while also breaking the symmetry of the dispersive bands, but do not affect the defect state, as shown in Fig.~\ref{fig1}(e). Without further modification, this defect would be localized around the top right corner of the system;
as illustrated in Fig.~\ref{fig1}(a), the defect state can then be moved along the edges and into the bulk by creating dimerization line defects separating regions where the role of $u$ and $u'$ or $v$ and $v'$ are interchanged.

\begin{figure}[t]
\includegraphics[width=\columnwidth]{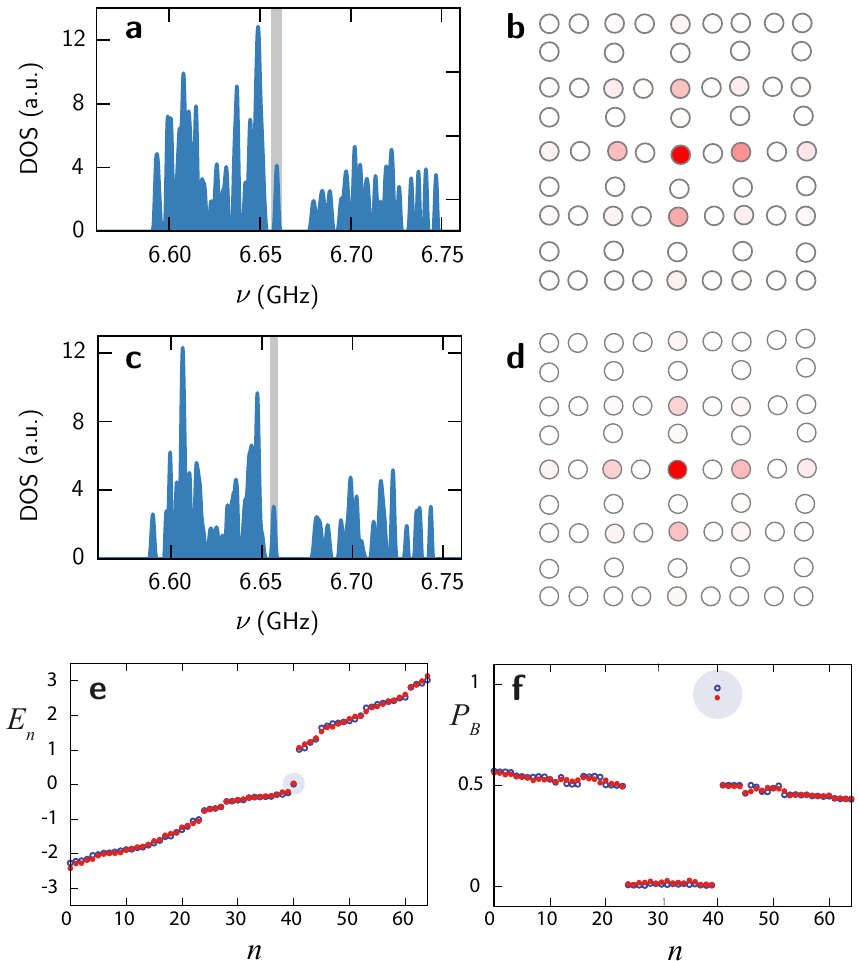}
\caption{\label{fig3} Results for disordered dimerized Lieb lattices with a central defect state.  (a,c) Experimental density of states, with the point-defect state shaded in light gray. (b,d) Experimental spatial distribution of the zero mode. In (a,b) the disorder fulfills the condition (\ref{eq:disorderconstraint}) while in (c,d) the disorder is generic. In both experiments the disorder strength is $W=0.175$. (e,f) Numerical results for generic disorder ($W=0.1$ blue circles, $W=0.25$ red dots) with couplings of the pristine system given by $u=v=4/3$, $u'=v'=2/3$, $w=0.4$, $w'=w'''=0.2$, $w''=0.1$  \cite{NumericalData}. Panel (e) shows the energy levels $E_n$ in ascending order, while panel (f) shows the weights $P_B$ of the eigenstates on the B sublattice (the corresponding sublattice polarization is $\mathcal{P}=1-2P_B$). The defect state is again shaded in light gray.
}
\end{figure}

Figure~\ref{fig2} shows the experimental results for configurations where the point defect sits at the corner [Fig.~\ref{fig2}(a-c)] or at the center [Fig.~\ref{fig2}(d-f)] of the system.
Fig.~\ref{fig2}(a) and (d) show the corresponding density of states (DOS). In agreement with the scenario proposed in this paper, the next-nearest neighbor couplings break the symmetry of the extended bands and isolate the zero mode (light gray zone) from the original flat band (dark gray zone), which now spreads over the whole system but remains mainly confined to the A sublattice  [Fig.~\ref{fig2}(c), (f)]. As expected, the state associated with the spectrally isolated peak displays a spatially localized profile with intensity confined to the B sublattice [Fig.~\ref{fig2}(b),(e)].

\subsection{Robustness against symmetry-preserving and generic disorder}

By construction, the created defect state is insensitive to any disorder in the couplings $w,w',w'',w'''$. Disorder in the couplings $u,u',v,v'$ modifies the wavefunction, but does not affect its energy or sublattice polarization as long as each plaquette fulfills the constraint
\begin{equation}
\label{eq:disorderconstraint}
u_{nm}v'_{nm}u'_{nm+1}v_{n+1m}=v_{nm}u'_{nm}u_{nm+1}v'_{n+1m},
\end{equation}
with couplings enumerated by the unit-cell index of the corresponding A sites [see Fig.~\ref{fig1}(a)].
In the continuum limit, where the displacement $\mathcal{A}$ of the virtual Dirac point can be interpreted as an imaginary vectorpotential, this condition amounts to a vanishing pseudomagnetic field induced by the deformation, $\nabla\times \mathcal{A}=0$.

Figure~\ref{fig3}(a) shows the experimental DOS of a dimerized Lieb lattice with a central defect and  disorder that respects the constraint \eqref{eq:disorderconstraint}.
The disordered positions were generated on the level of the tight-binding model, where the couplings
$u_{nm}'\in [(1-W)u',(1+W)u']$, $v_{nm}\in [(1-W)v,(1+W)v]$, $v_{nm}'\in [(1-W)v',(1+W)v']$ were chosen randomly from box distributions with $W=0.175$, while the couplings $u_{nm}$ were determined from the constraint \eqref{eq:disorderconstraint}. These couplings were then translated into a compatible configuration of random displacements, which also introduces disorder into the couplings $w^{(i)}$.
The disks were then placed accordingly.
In comparison with Fig.~\ref{fig2}(d), we see that the disorder is sufficiently strong to mix the extended states, but does not affect the energy of the zero mode, which remains spectrally isolated. Furthermore, the spatial distribution of this mode shown in Fig.~\ref{fig3}(b) exhibits the same localization as observed for the nondisordered lattice in Fig.~\ref{fig2}(e).

Encouragingly for practical applications, the defect state remains well isolated and localized also for generic disorder. In Fig.~\ref{fig3}(c,d), this is confirmed experimentally for disorder of the same strength $W=0.175$, obtained by generating displaced positions of the disks as before but ignoring the constraint \eqref{eq:disorderconstraint}.
As such disorder is still correlated (displacing a disk modifies several couplings systematically), we also carried out numerical calculations in which all couplings were randomly perturbed. As shown in
Fig.~\ref{fig3}(e), the spectral isolation of the defect state remains intact even at larger values of disorder; furthermore, the defect state remains predominantly localized on the B sublattice, as shown in Fig.~\ref{fig3}(f).

This robustness can be understood as an added benefit from the spectral isolation and sublattice polarization of the defect state, as this automatically suppresses its sensitivity to generic coupling disorder in a perturbative treatment. Due to the sublattice polarization, the disorder only contributes from the second order, where energy levels further repel. Condition \eqref{eq:disorderconstraint} leaves only one effective disorder freedom per plaquette, so that the relevant matrix elements only become comparable to the spectral isolation energy $\Delta_0$ when the disorder is sizeable ($W\gtrsim 0.5$ under the conditions investigated here). An equivalent amount of onsite disorder, which breaks the original chiral symmetry, has a stronger effect on the spectral isolation, but only weakly affects the sublattice polarization as the matrix elements relevant for the leading order again vanish. The latter type of disorder is small in all experiments.

\subsection{Mode selection}
\begin{figure}
\includegraphics[width=\columnwidth]{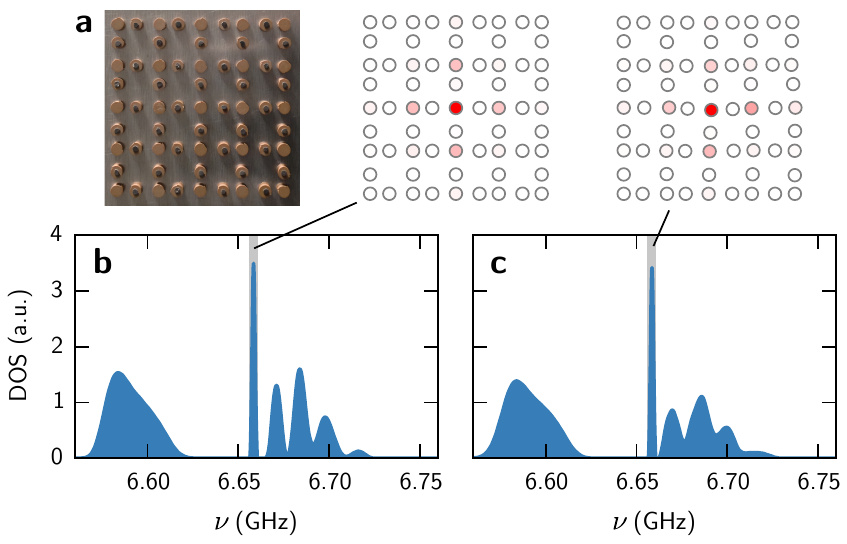}[t]
\caption{\label{fig4} (a) Picture of the experimental set-up realizing a dimerized Lieb lattice with identical absorption on the A sites, again configured to support a
defect state at the center. (b) Density of state of the regular system. (c) Density of state of the disordered system [same configuration as in Fig.~\ref{fig3}]. The absorption suppresses the flat-band and extended states, but does not suppress the defect state, which remains localized (insets).
}
\end{figure}

The partial chiral symmetry remains intact if one includes a possibly inhomogeneous onsite potential on the A sublattice. This includes the choice of an imaginary onsite potential that corresponds to absorption, in  generalization of non-hermitian $PT$-symmetric optics \cite{Guo09,Rut10,Reg12,Fen13,Eic13,Fen14,Hod14}. In absence of the next-nearest neighbour couplings, the extended states remain sublattice-balanced for $\gamma<2\Delta$ and decay according to  ${\rm Im}\, E= -\gamma/2$; the states from the flat band decay twice as fast (${\rm Im}\, E=-\gamma$), while the point-defect state stays pinned at $E=0$ and thus remains unaffected. In the presence of additional amplification, this corresponds to a topological mechanism of mode selection for lasing \cite{Sch13a,Sch13b}. In our experiments, we implement the absorption by placing elastomer patches on the top of each A site \cite{Pol15}, thus inducing a uniform absorption on the A sublattice. The selection mechanism can be clearly seen in Fig.~\ref{fig4} where two situations are investigated. Fig.~\ref{fig4}(a,b) shows the effect of the absorption on a dimerized Lieb lattice with central defect [same configuration as in Fig.~\ref{fig2}(d)], while Fig.~\ref{fig4}(c) corresponds to the disordered system shown in Fig.~\ref{fig3}(a,b). In both cases, the original flat band and the two dispersive bands are considerably suppressed, while the zero mode is not affected both spectrally as well as  in its spatial distribution. Therefore, absorption on the A sublattice can be used to further enhance the spectral isolation of the defect state.

\section{Conclusions}

In summary, we showed that a partial breaking of chiral symmetry can be employed to stabilize and isolate point-defect zero modes in two-dimensional bipartite systems with flat bands.
Remarkably, the point-defect zero mode resides on the sublattice with the smaller number of sites (the minority sublattice), and thus displays a sublattice polarization with an opposite sign to the flat-band states, which  reside on the majority sublattice.
The flat band states can then be removed by breaking the symmetry on the majority sublattice, without any effects on the point-defect zero mode.
As we demonstrated experimentally in a microwave realization of a dimerized Lieb lattice with next-nearest neighbour couplings, the remaining zero mode is spectrally isolated and spatially localized. These features increase the robustness of the state, which in this setting is useful for applications in photonic mode shaping and guiding, including for arrays of coupled waveguides  where lattices are designed to govern the propagation dynamics. The sublattice polarization also provides a route to mode selection via loss imbalance, which can be extended to laser settings where they simplify the mode competition. In artificial two-dimensional flat-band materials, the partial symmetry-breaking described here provides a practical mechanism to design robust defect states with a unique mode profile and controllable spectral isolation.

\section{Acknowledgments}

This research was supported by EPSRC via Grant No. EP/J019585/1 and Programme Grant  No. EP/N031776/1.
MB, UK and FM acknowledge helpful contributions from Ioannis Pitsios during the early stage of this work.
The numerical data are openly available \cite{NumericalData}.

\appendix

\section{Conventional and partial chiral symmetry in finite systems}

In the main text we focus on  Bloch Hamiltonians and take care of boundary conditions by exploiting the anomalous sublattice polarization of the evanescent zero-mode Bloch waves.
To prepare a more detailed discussion of the Lieb lattice, we first describe the notions of a conventional chiral and partial chiral symmetry in the context of general (possibly finite and non-periodic) bipartite lattices.

Such systems still consist of two sublattices (A sites and B sites). The chiral symmetry is realized when these sublattices are coupled together to result in an off-diagonal Hamiltonian,  corresponding to real-space tight-binding equations
\begin{equation}
E\psi=H\psi,\quad H=\left(
\begin{array}{cc}0 & T_{AB} \\ T_{BA} & 0 \\ \end{array}\right),  \quad
\psi=\left(
\begin{array}{c}\psi_A \\ \psi_B \\ \end{array}\right).
\label{eq:tb2}
\end{equation}
Here $T_{AB}=T_{BA}^\dagger$ describes the coupling of the A sites to the B sites, whose amplitudes are collected in the vectors $\psi_A$ and $\psi_B$. The Hamiltonian possesses a chiral symmetry, $\tau_zH\tau_z=-H$, where the Pauli matrix $\tau_z$ acts in sublattice space.

In a finite system with $N_A$ A sites and $N_B$ B sites, $T_{AB}$ is an $N_A\times N_B$-dimensional matrix. We set $N_A\geq N_B$ and call the A and B sites the majority and minority sublattice, respectively. Generically, the system then possesses $N_A-N_B$ sublattice-polarized zero modes, given by the solutions of the under-determined linear system  $T_{BA}\psi_A=0$ while $\psi_B=0$ \cite{sutherland,Lie89}. The chiral symmetry enforces that the remaining $2N_B$ states occur in pairs with energy $E$ and $-E$, and furthermore all possess equal weight  on both  sublattices, $|\psi_A|^2=|\psi_B|^2=1/2$, hence a vanishing sublattice polarization $|\psi_A|^2-|\psi_B|^2=\psi^\dagger \tau_z\psi$. The states in each pair are related by $\tau_z$, corresponding to a sign change of the amplitude on the minority sublattice. Their vanishing sublattice polarization  then follows from the identity
\begin{equation}
\psi^\dagger (H\tau_z)\psi=-\psi^\dagger (\tau_zH)\psi \Rightarrow E\psi^\dagger \tau_z\psi= -E\psi^\dagger \tau_z\psi.
\end{equation}

As we have shown in the main text, the presence of real or virtual Dirac points allows to increase the zero-mode count by two, with one mode $\psi_+$ supported by the A sublattice and the other mode $\psi_-$  supported by the B sublattice. This occurs in the general setting when columns of  $T_{AB}$ are linearly dependent of each other.
The degeneracy of these modes can now again be lifted by introducing terms $T_{AA}$ into the Hamiltonian,
\begin{equation}
H=\left(
\begin{array}{cc}T_{AA} & T_{AB} \\ T_{BA} & 0 \\ \end{array}\right).
\end{equation}
This reduces the chiral symmetry to
\begin{equation}
[\tau_zH\tau_z]_{BB}=[-H]_{BB},
\end{equation}
which we again can take as the definition of a partial chiral symmetry. This modification does not affect both the energy as well as the wavefunction of the zero mode $\psi_-$ on the B sublattice, but generically affects all other states---the finite-energy states just as much as the zero modes localized on the A sublattice, including $\psi_+$.

\section{Lieb lattice}

\begin{figure*}[t]
\includegraphics[width=0.95\linewidth]{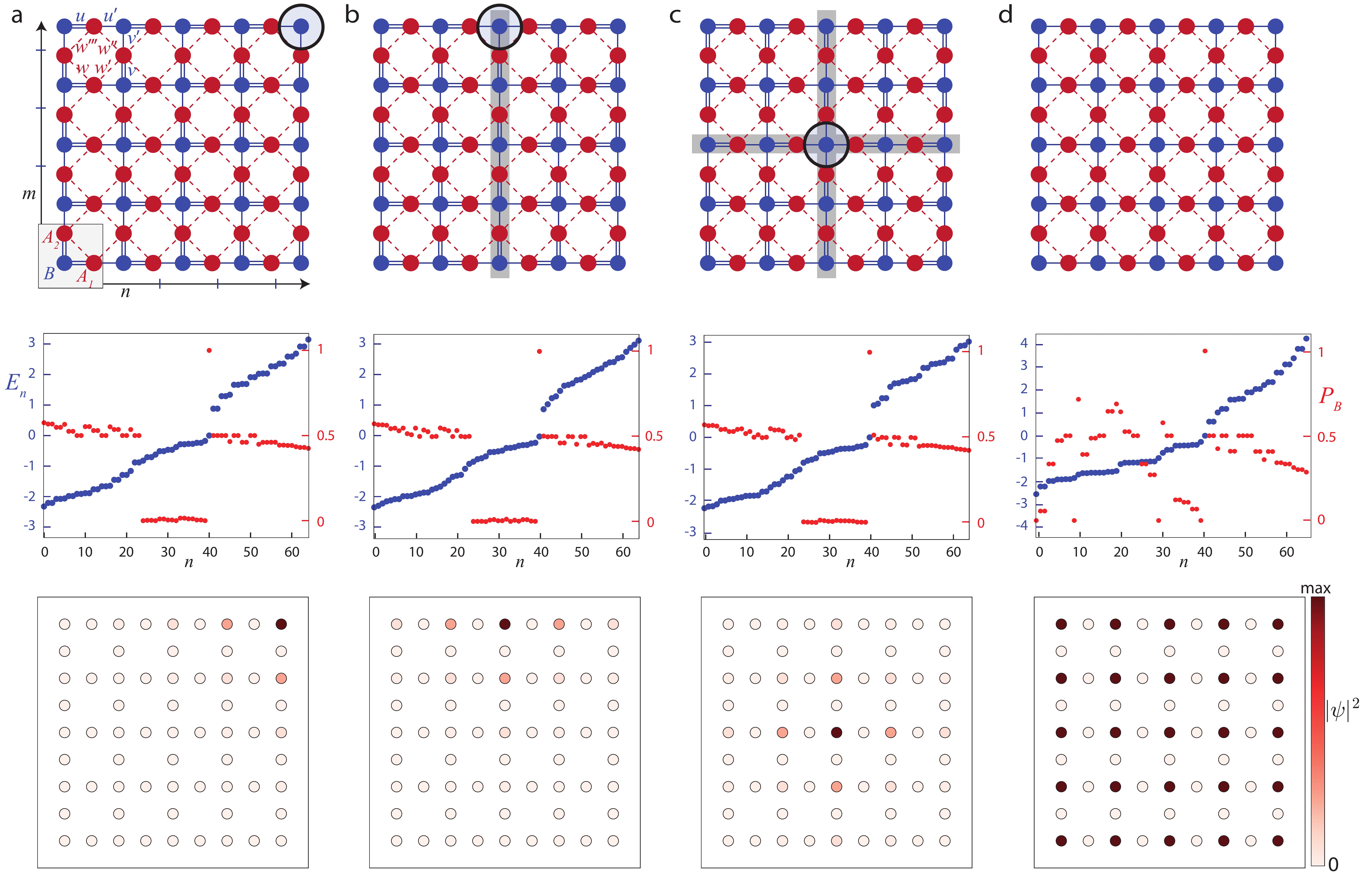}
\caption{\label{fig5} (a) The top panel shows a finite Lieb lattice with dimerized couplings $u$, $u'$, $v$, $v'$, as well as next-nearest-neighbour couplings $w$, $w'$, $w''$, $w'''$ that reduce the chiral symmetry.
The circle highlights the site in the top right corner, around which the defect state is localized if $u>u'$ and $v>v'$ (the site is then only coupled by the weak links).
The  middle row shows the energy levels $E_n$ (blue filled circles), in ascending order, and the weights $P_B$ (red filled circles) of the eigenstates on the B sublattice, for $u=v=4/3$, $u'=v'=2/3$, $w=0.4$, $w'=w'''=0.2$, $w''=0.1$. The bottom panel shows the spatial distribution of the isolated zero mode, which is only supported by the B sublattice ($P_B=1$).
(b) The lattice with a vertical dimerization line-defect (gray). Now the defect state is localized around the circled site at the top edge, where again all couplings are weak. (c) The lattice with two dimerization line-defects (gray) that cross in the center of the system, again resulting in a site that only has weak couplings.
The defect state is now localized around this central site.
(d) Results for the Lieb lattice without dimerization, with $u=v=u'=v'=1$ and $w=w'=w''=w'''=0.5$. The zero mode is now extended, and more fragile to hybridization with the other states in the system.}
\end{figure*}

We here give a detailed discussion of the formation of the defect state in the dimerized Lieb lattice, first in the tight-binding description and then in the continuum limit.
\subsection{Tight-binding description}
The dimerized Lieb lattice with nearest-neighbour couplings is reproduced in Fig.~\ref{fig5}.
In the tight-binding description, each site $X=A_1$, $A_2$ or $B$ in a given unit cell provides a
basis state $|X_{n,m}\rangle$, with the cells enumerated by a pair of integers $n,m$. The tight-binding Hamiltonian is then given by
\begin{eqnarray}\label{tb2DSSH}
 H=\sum_{nm}&&\Big[\Big(u_{nm}|B_{n,m}\rangle + u'_{nm}|B_{n+1,m}\rangle  \Big)\langle A_{1;n,m}|
 \nonumber
 \\
 &&+
  \Big(v_{nm}|B_{n,m}\rangle + v'_{nm}|B_{n,m+1}\rangle  \Big)\langle A_{2;n,m}|
 \nonumber
 \\
 &&+
  \Big(w_{nm}|A_{2;n,m}\rangle + w'_{nm}|A_{2;n+1,m}\rangle
   \nonumber
 \\
 &&+w''_{nm}|A_{2;n+1,m-1}\rangle + w'''_{nm}|A_{2;n,m-1}\rangle\Big) \nonumber
 \\ &&\times \langle A_{1;n,m}|\Big]+
   h.c.
\end{eqnarray}

For the infinitely periodic system with $u_{nm}=u$ etc., we can seek solutions in the form of a Bloch wave
\begin{equation}
|\psi(\mathbf{k})\rangle=\sum_{X=A_1,A_2,B}\sum_{nm}e^{i\mathbf{k}\cdot(n,m)}\varphi_X(\mathbf{k})|X_{n,m}\rangle,
\end{equation}
which results in the Bloch eigenvalue problem $E\varphi=h\varphi$,
\begin{eqnarray}
\label{eq:liebbloch}
h(\mathbf{k})&=&
U(k_x)|A_1\rangle\langle B|+V(k_y)|A_2\rangle\langle B|+W(\mathbf{k})| A_1\rangle\langle A_2|
\nonumber\\
&&+h.c.,
\nonumber\\
|\varphi(\mathbf{k})\rangle&=&\varphi_{A_1}(\mathbf{k})|A_1\rangle+\varphi_{A_2}(\mathbf{k})|A_2\rangle+\varphi_{B}(\mathbf{k})|B\rangle ,
\nonumber \\
U(k_x)&=& u+u'e^{ik_x},\quad V(k_y)= v+v'e^{ik_y} ,
\nonumber \\
W({\bf k})&=&w+w'e^{ik_x}+w''e^{ik_x-ik_y}+w'''e^{-ik_y},
\end{eqnarray}
where we set the lattice constant $a\equiv 1$.

For the well-studied case of identical nearest-neighbour couplings $u=u'=v=v'$ and $w^{(i)}=0$,
the corresponding band structure consists of a flat band of states $|\varphi_0(\mathbf{k})\rangle \propto V(-k_y)|A_1\rangle-U(-k_x)|A_2\rangle$ localized on the A sublattice, and a conical dispersion relation
\begin{equation}
\label{eq:liebdisp}
E^2=|U(k_x)|^2+|V(k_y)|^2
\end{equation}
of extended states, with a Dirac point at the M point ${\bf K}_0=(\pi,\pi)$ in the corner of the Brillouin zone. The two degenerate extended states at the Dirac point can be combined into a sublattice-polarized states $|\varphi_+\rangle$, localized on the A sublattice, and $|\varphi_-\rangle=|B\rangle$, where the latter corresponds to a real-space wavefunction with $|\psi_-\rangle=\sum_{nm}(-1)^{nm}|B_{n,m}\rangle$. This state is compatible with the boundary conditions in finite systems terminated on the B sublattice, as used in the main text and also shown in Fig.~\ref{fig5}.

To see how this state becomes confined by a dimerization pattern (alternating couplings $u$, $u'$ along the $x$ direction and $v$, $v'$ along the $y$ direction), we first confirm from Eq.~\eqref{eq:liebdisp} that this opens a gap $2\Delta$ for the extended states, with $\Delta^2 =E^2(\mathbf{K}_0)=(u-u')^2+(v-v')^2$.
The state $|\varphi_{-}\rangle$ still survives as a state with a complex wave number $\mathbf{k}_\pm=\mathbf{K}_0\pm i(\ln u/u',\ln v/v')$, according to the analytical continuation of the Dirac point into the complex plane. The corresponding  real-space wavefunction
\begin{equation}
\label{eq:psiminus}
|\psi_-\rangle =\sum_{nm}(-u/u')^n(-v/v')^m|B_{n,m}\rangle
\end{equation}
still vanishes on the A sublattice,  while
on the B sublattice it now has an exponentially varying envelope.
In a rectangular finite system, this state remains compatible with the boundary conditions if the system terminates on the B sublattice, and then localizes around a corner of the system. We choose an orientation so that this is the top right corner (this corresponds to $u>u'>0$, $v>v'>0$ and can always be achieved by a rotation of the system and redefinition of the unit cell), as indicated in Fig.~\ref{fig5}(a).

To confirm that the state $|\psi_-\rangle$ can  be moved along the edges and into the bulk, we consider line defects separating regions where the role of $u$ and $u'$ or $v$ and $v'$
are interchanged, see Fig.~\ref{fig5}(b,c). Along a vertical  line defect, we encounter $A$ sites that are coupled by $u'$ both to the right and to the left, while along a horizontal line defect  they are coupled by $v'$ both to the bottom and to the top.
We denote by $n=m=0$ the crossing point of two such  line defects, or the point where a single  line defect meets the upper or right boundary. This results in a point-defect state with real-space wavefunction
\begin{equation}
\label{eq:pointdefectstate}
|\psi_-\rangle=\sum_{nm}(-u/u')^{|n|}(-v/v')^{|m|}|B_{n,m}\rangle,
\end{equation}
which now is exponentially localized into all directions.
Arbitrary next-to-nearest neighbour couplings $w^{(i)}$ between A$_1$ and A$_2$ sites break the symmetry of the extended bands and introduce a dispersion to the flat band, which moves away to finite energies, as illustrated in Fig. 1 of the main text. However, as a consequence of the partial chiral symmetry, the energy and wavefunction of  the point-defect state \eqref{eq:pointdefectstate} remains unchanged.

These statements are verified by  the numerical results in the middle and bottom row of Fig.~\ref{fig5}. In each case, we find a spectrally isolated zero mode that is localized on the B sublattice, and exhibits the expected spatial profile \eqref{eq:pointdefectstate}.

\begin{figure}[t!]
\includegraphics[width=\columnwidth]{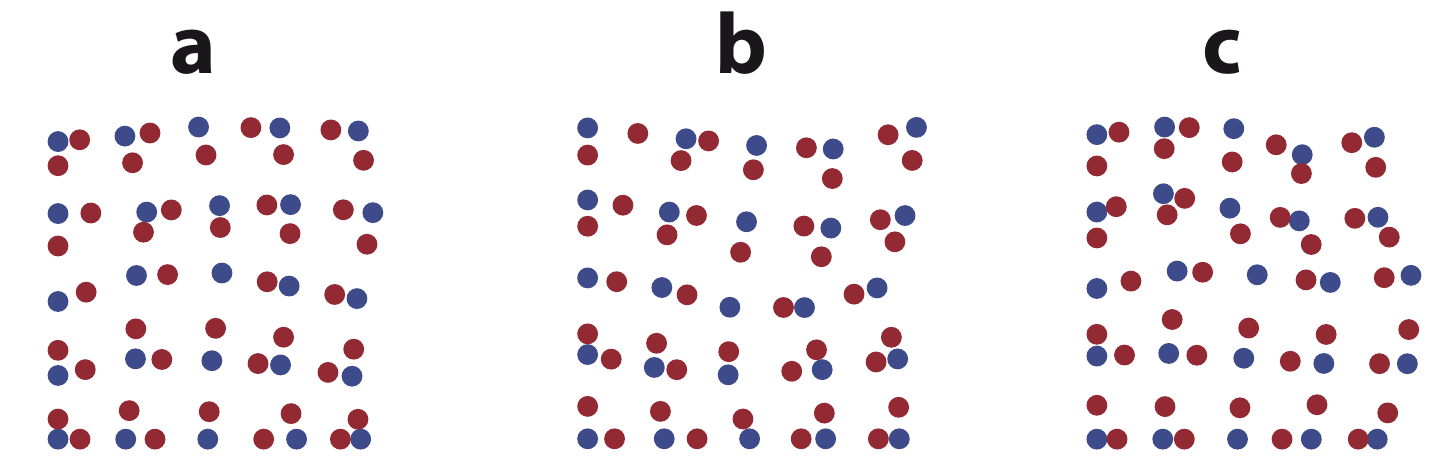}
\caption{Disordered Lieb lattices in presence of disorder with strength $W=0.25$, represented in real space where larger distances $d$ correspond to weaker couplings $t$. As for the experiment we use an exponential distance-dependence $t\propto\exp(-d/d_0)$, but choose $d_0$ to make the disorder more visible.
Panels (a,b) show configurations that obey the constraint \eqref{eq:disorderconstraint} (generated as described in the main text), while panel (c) shows a configuration with unconstrained disorder.
}\label{fig6}
\end{figure}

By construction, the point-defect state  \eqref{eq:pointdefectstate} is insensitive to any disorder in the couplings $w^{(i)}_{nm}$. From the condition $T_{AB}\psi_B=0$, we can further verify that disorder in the couplings $u_{nm},u'_{nm},v_{nm},v'_{nm}$ does not affect the energy or sublattice polarization of the state as long as each plaquette fulfills the constraint \eqref{eq:disorderconstraint}.
Compatible types of disorder include quasi-one-dimensional disorder in which the hoppings $u_{nm},u'_{nm}$ only depend on the unit-cell index $n$ while the hoppings $v_{nm},v'_{nm}$ only depend on the unit-cell index $m$. As described in the main text, the constraint also allows for much richer disorder configurations, two of which are illustrated in Fig.~\ref{fig6}(a,b), while Fig.~\ref{fig6}(c) shows a case of generic disorder.

While we focussed on the implementation of the point-defect state in a photonic setting,
we remark that is also interesting to consider this system in an electronic context. In the ground state the single-particle states are then occupied according to the Pauli principle up to a Fermi energy $E_F$.
From a simple counting argument \cite{Ser08}, we then find that
a uniform charge density at finite filling requires a half-population of the point-defect state.
At Fermi energy  $E_F=0^{\pm}$, the state thus serves as a fractional charge $\pm 1/2$ against
a uniform background density on the minority sublattice. This resembles the situation for  a $Z_4$ vortex in a dimerized square lattice (which does not display a flat band)
\cite{Ser08,Ser08a}.

\subsection{Topological characterisation in the continuum limit}

An important paradigm of a dimerized system with topologically protected point-defect zero modes is the Su-Schrieffer Heeger (SSH) model \cite{Su79}, which consists of a one-dimensional tight-binding chain with alternating couplings
\begin{equation}
H=\sum_n(u|A_{n}\rangle+u'|A_{n-1}\rangle)\langle B_n|+h.c.
\end{equation}
Here we give an interpretation of the dimerized Lieb lattice as a two-dimensional generalization of the SSH model.
This connection is most usefully established in the continuum limit, which sheds further light on the topological features of the system.

For the SSH model, the continuum limit is obtained by a gauge transformation $|A\rangle \rightarrow e^{ik_x/2}|A\rangle$, followed by a gradient expansion in $k_x\rightarrow\pi-i\partial_x$. This results in a massive Dirac Hamiltonian, the Jackiw-Rebbi model \cite{Jac76}  with
\begin{equation}
E\psi=H\psi,\quad
H=iv_{0}\sigma_x\partial_x-m\sigma_y,\quad\psi=\left(
\begin{array}{c}\psi_A(x)  \\ \psi_B(x) \end{array}\right),
\end{equation}
where $m=u'-u$ and $v_{0}=(u+u')/2$.
For constant mass $m$, the dispersion relation is given by $E^2=m^2+v_0^2 k_x^2$. When two regions of opposite mass are joined together, $m(x)=M\,\mathrm{sgn}\,x$ with $M>0$, one finds an exponentially localized defect mode of the form $\psi_A(x)=0$, $\psi_B(x)\propto\exp(-|x|M/v_0)$.

For the dimerized Lieb lattice, we obtain the continuum limit from the Bloch Hamiltonian \eqref{eq:liebbloch}
by performing the gauge transformation $|A_1\rangle\to e^{ik_x/2}|A_1\rangle$, $|A_2\rangle\to e^{ik_y/2}|A_2\rangle$
and expanding around $\mathbf{k}\rightarrow \mathbf{K}_0-i\nabla$.
We then obtain a massive Dirac Hamiltonian
\begin{align}\label{dirac}
\qquad&\mathcal{H}     &\hspace{-1.5em}&= \mathcal{H}_X+\mathcal{H}_Y+\mathcal{H}_{XY} , \\
\qquad&\mathcal{H}_X   &\hspace{-1.5em}&= iv_X\lambda_4\partial_x-m_X\lambda_5,  \\
\qquad&\mathcal{H}_Y   &\hspace{-1.5em}&= iv_Y\lambda_6\partial_y-m_Y\lambda_7, \\
\qquad&\mathcal{H}_{XY}&\hspace{-1.5em}&= m_{XY}\lambda_1 +i\lambda_2(\tilde v_X\partial_x +\tilde v_Y\partial_y ),
\end{align}
where $\lambda_i$ are the Gell-Mann matrices.
The effective velocities are $v_X=(u+u')/2$ and $v_Y=(v+v')/2$, as well as
$\tilde v_X=(-w-w'+w''+w''')/2$ and $\tilde v_Y=(w-w'-w''+w''')/2$,
while the masses are given by $m_X=u'-u$, $m_Y=v'-v$, $m_{XY}=w-w'+w''-w'''$.

This model recovers the low-energy part of the band structure.
In absence of the couplings $w^{(i)}$, we obtain a flat band of zero modes on the majority sublattice, and dispersive bands with
$E^2=m_X^2+m_Y^2+v_X^2k_x^2+v_Y^2k_Y^2$. For the system with a vertical dimerization line-defect as in Fig.~\ref{fig5}(b), $m_X(x)=M_X\,\mathrm{sgn}\,x$ with $M_X=u-u'>0$, so that the sign of the mass changes at the interface. This creates a conduction channel within the gap of extended states, described by  exponentially confined modes
\begin{align}
|\psi\rangle&\propto e^{-\frac{M_X}{v_X} |x|}e^{-i k_y y}\nonumber \\
&\times
\left[|B(x,y)\rangle+ \frac{im_Y-v_Yk_y}{E} |A_2(x,y)\rangle\right],
\end{align}
which form
two symmetric bands with $E^2 =m_Y^2+v_Y^2k_y^2$.
The channel is thus described by a Jackiw-Rebbi model for the transport along $y$.

Within this channel, an additional line defect parallel to the $x$ axis should therefore produce a point-defect zero-mode that decays exponentially in all directions.
We can verify this directly. For a central dimerization defect as in Fig.~\ref{fig5} (c),
the effective masses are given by $m_X(x)=M_X\,\mathrm{sgn}\,x$ and $m_Y(y)=M_Y\,\mathrm{sgn}\,y$, with $M_X=u-u'>0$ and $M_Y=v-v'>0$.
While $[\mathcal{H}_X, \mathcal{H}_Y]\neq 0$, we have $([\mathcal{H}_X, \mathcal{H}_Y])_{BB}=0$, so that the point-defect zero mode on the minority sublattice can be obtained by separation of variables,
\begin{equation}\label{zm_dirac}
|\psi_{-}(x,y)\rangle\propto e^{-\frac{M_X}{v_X} |x|}e^{-\frac{M_Y}{v_Y} |y|}|B(x,y)\rangle .
\end{equation}
This indeed corresponds to the continuum limit of the tight-binding solution \eqref{eq:pointdefectstate}.
We can again verify directly that this state is not affected by the symmetry-breaking terms $H_{XY}$.

Thus, within the continuum limit the point-defect state can be associated with a nontrivial background mass pattern in the system.
Note that when this mass pattern is smoothed out, the mass-gap order parameter $m_X(\mathbf{r})+im_Y(\mathbf{r})$ has a non-trivial topology, with a finite winding number as one encircles the origin along a closed loop. This observation establishes an additional connection to systems with charge fractionalization \cite{Hou07}.

We conclude by identifying the continuum interpretation of the constraint \eqref{eq:disorderconstraint}. In the tight-binding model, the position of the virtual
Dirac point can be written as
$\mathbf{k}_\pm=\mathbf{K}_0\pm i\mathcal{A}$, where
the displacement $\mathcal{A}=[\ln(u/u'),\ln(v/v')]$ can be interpreted as an imaginary pseudo vector potential (in analogy to the description of strained graphene). We then can rewrite Eq.~\eqref{eq:disorderconstraint} as
\begin{equation}
\mathcal{A}_{y;n+1,m}-\mathcal{A}_{y;n,m}=\mathcal{A}_{x;n,m+1}-\mathcal{A}_{x;n,m}
\rightarrow \partial_x \mathcal{A}_y=\partial_y \mathcal{A}_x,
\end{equation}
therefore $\nabla\times \mathcal{A}=0$.
Thus, the constraint on the disorder can be naturally interpreted as the condition of a vanishing deformation-induced pseudomagnetic field.


\begin{thebibliography}{35}
\bibitem{fn1} In asuitable basis, a conventional time-reversal symmetry $\mathcal{T}$ can be represented as a complex conjugation, and squares to $\mathcal{T}^2=1$. This is the typical case for optical systems, while time-reversal symmetries with $\mathcal{T}^2=-1$ can be encoutered in electronic systems with broken spin-rotation symmetry.
\end{thebibliography}
\end{document}